# Heterodyne detection of radio-frequency electric fields using point defects in silicon carbide

Gary Wolfowicz[1], Christopher P. Anderson[1,2], Samuel J. Whiteley[1,2], David D. Awschalom[1,3,*]

[1]Institute for Molecular Engineering, University of Chicago, Chicago, Illinois 60637, USA
[2]Department of Physics, University of Chicago, Chicago, Illinois 60637, USA
[3]Institute for Molecular Engineering and Materials Science Division, Argonne National Laboratory, Lemont, Illinois 60439, USA

[*]Corresponding Author: awsch@uchicago.edu

**Sensing electric fields with high sensitivity, high spatial resolution and at radio frequencies can be challenging to realize. Recently, point defects in silicon carbide have shown their ability to measure local electric fields by optical charge conversion of their charge state. Here we report the combination of heterodyne detection with charge-based electric field sensing, solving many of the previous limitations of this technique. Owing to the non-linear response of the charge conversion to electric fields, the application of a separate "pump" electric field results in a detection sensitivity as low as $1.1\ (\mathrm{V/cm})/\sqrt{\mathrm{Hz}}$, with near-diffraction limited spatial resolution and tunable control of the sensor dynamic range. In addition, we show both incoherent and coherent heterodyne detection, allowing measurements of either unknown random fields or synchronized fields with higher sensitivities. Finally, we demonstrate in-plane vector measurements of the electric field by combining orthogonal pump electric fields. Overall, this work establishes charge-based measurements as highly relevant for solid-state defect sensing.**

Optically-active point defects in solid-state materials have emerged as a novel class of universal local sensors, able to detect magnetic field, electric field, temperature and strain with high spatial resolution[1–6]. These characteristics, initially developed using the nitrogen-vacancy center in diamond, are applicable to a variety of defects in many host materials[7–10] with different sensing capabilities depending on their spin, optical and charge states[11–13].

The long electron spin lifetime and coherence desired for spin-based detection require low sensitivity to strain and electric fields such that the spin does not interact with its solid-state host, e.g. from spin-orbit interactions or effects from nearby charges[14]. Spin sensing is therefore not well suited for sensing electric fields, both by choice to be robust against the environment and simply because the spin is a magnetic moment. Instead, the orbital and charge states of defects appear more relevant as they are directly perturbed by the local electric environment[10].

Recently, it has been shown that the charge stability of defects can be used to measure radio-frequency electric fields using divacancies and silicon vacancies in silicon carbide (SiC)[11]. This technique, called Electrometry by Optical Charge Conversion (EOCC), consists of optically controlling the defect's charge state between two configurations with different photoluminescence intensity. The rate of photo-induced charge conversion is sensitive to electric fields, likely due to changes in the various ionization and capture cross-sections of the defect[11]. EOCC exhibits a non-linear (near-quadratic) response to the electric field, very similar to that of a power photo-detector in optics or a mixer in electronics. By analogy, this implies that the variety of heterodyne techniques[15] developed in these systems applies equally to EOCC, enabling improvements in sensitivity and a range of additional tools for sensing radio-frequency electric fields.

In this work, we first present how to employ heterodyne detection with EOCC, both in the incoherent and coherent regime. Second, by applying and controlling a pump electric field, we achieve high-resolution radio-frequency spectroscopy, gain and dynamic range control, linearization and increased sensitivity to electric fields. We finally present a method to obtain vector information of the sensed field.

We perform experiments on divacancy ensembles in 4H-SiC at 20 K using a confocal microscopy setup. The sample is a 20 µm epitaxial semi-insulating layer grown on top of a semi-insulating substrate from Norstel AB. The defects are created by carbon implantation at a 7° tilt with a $1 \times 10^{12}$ cm$^{-2}$ dose at 170 keV followed by annealing at 850°C in argon. The divacancies, in their neutral charge state, emit near-infrared (1100-1300 nm) photoluminescence after excitation with a 976 nm laser. This laser also converts the charge state of the defects from neutral (bright) to negative (dark, no photoluminescence)[16,17]. A low power 365 nm light-emitting diode re-pumps the charge state to neutral. Since only the neutral state is bright, the photoluminescence provides a direct measurement of the averaged charge state of the divacancy ensemble. A four-gate device is fabricated on top of the sample using electron-beam lithography and 30 nm Ti/150 nm Au are deposited with electron-beam evaporation, enabling electric fields to be applied in two orthogonal in-plane directions as shown in Fig. 1(a). Each opposite pair of gates are connected to the outputs of a 180-degree splitter for differential voltages, with input from a radio-frequency signal generator. When unused, the gates are grounded. All values of electric field are estimated at the center of device ($X = Y = 0$ µm) from the experimental input voltages at the gates and finite-element modeling (COMSOL Multiphysics®).

The EOCC measurements are obtained by measuring the average charge state of the divacancy, via the ensemble photoluminescence (*PL*), under steady-state illumination at 976 nm and 365 nm and in the presence of the radio-frequency electric field. In the case of standard (non-heterodyne) EOCC, the electric field $E$ is periodically turned on and off for lock-in detection, resulting in a contrast value $(PL(E) - PL(E = 0))/PL(E = 0)$. In Fig. 1(b) we measure a two-dimensional map of the EOCC contrast (left), and plot the square root to obtain the electric field profile and compare to finite-element simulation (right). Near the metallic gates, the EOCC signal is less accurate due to saturation at high electric field values (~kV/cm) and modified laser excitation/light collection. A few micrometers away from the metallic gates, we have a good agreement with the model which allows us to calibrate the applied electric field value.

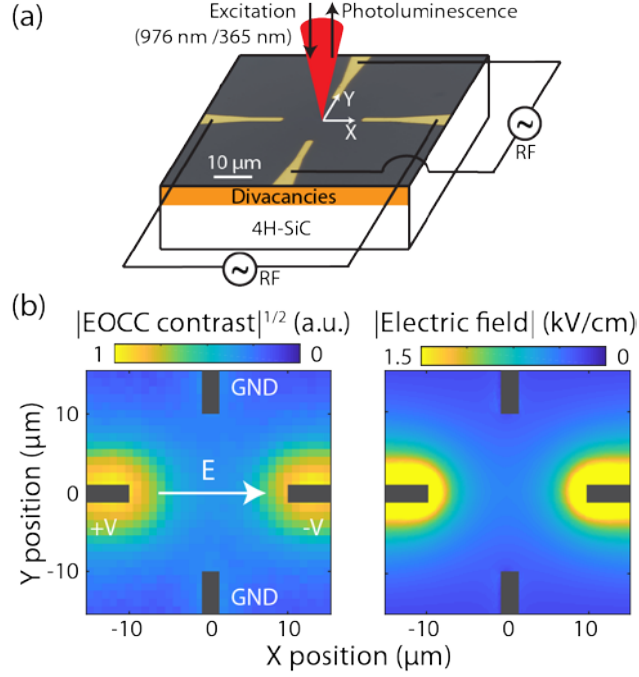

**Fig. 1. Electrometry device.** (a) Schematics of the sensing device with microscope image (top surface). Divacancies are implanted near the surface ($\approx 300\ nm$ deep) of the 4H-SiC semi-insulating substrate. Four equidistant electrodes are patterned and fabricated on top of the substrate to apply both pump and sensed radio-frequency (RF) electric fields. Optical excitations for both charge pumping and photoluminescence are focused near the surface, and the photoluminescence is collected into an objective. (b) Left: standard EOCC contrast for an in-plane radio-frequency (65 MHz) electric field $E$ (or voltages V). Right: corresponding electric field (static) estimated from finite-element modeling. The electric field is limited to 1.5 kV/cm near the electrodes for contrast. The black rectangles represent the metal electrodes.

The non-linear behavior of the EOCC response allows for heterodyne detection of the electric field. The principle is as follows: a pump electric field $E_{\text{pump}}\cos(2\pi f_{\text{pump}} t)\vec{u_p}$ is applied simultaneously with the electric field to be sensed $E_{\text{sensed}}\cos(2\pi f_{\text{sensed}} t + \varphi)\vec{u_s}$, where $\varphi$ is the relative phase with respect with the pump and $\vec{u_p}$ and $\vec{u_s}$ are normalized vectors. The EOCC signal is proportional to $\langle ||\vec{E}||^2 \rangle_t$, where $\vec{E}$ is the total electric field and $\langle\ \rangle_t$ is the time average for signals at frequencies higher than the photodetector bandwidth (1 kHz). The output signal therefore contains an oscillating component $E_{\text{pump}} E_{\text{sensed}} \cos(2\pi \Delta f t + \varphi)\vec{u_p}\cdot\vec{u_s}$, where $\Delta f = f_{\text{pump}} - f_{\text{sensed}}$ is referred to as the beat frequency. As shown in Fig. 2(a), this signal is demodulated using lock-in detection at a frequency $f_{\text{ref}}$. When $f_{\text{ref}}$ matches $\Delta f$, we obtain a signal that is simply proportional to the product $E_{\text{pump}} E_{\text{sensed}}$. If the phase $\varphi$ is random, i.e. no assumptions are made on the field to detect, the heterodyne measurement is incoherent with only the magnitude of the signal available. If the phase is not random, for example in a driven or triggered system, then the measurement is coherent and both magnitude and phase are directly available from the lock-in detection.

We initially consider the case of incoherent heterodyne with collinear pump and sensed electric fields. In Fig. 2(b), we measure the lock-in signal magnitude as a function of the beat frequency (with $f_{\text{pump}} = 68$ MHz, chosen to avoid unwanted device/setup resonances) and observe a peak at $f_{\text{ref}}$. Since the sensed electric field is artificially generated by a low-noise signal generator, the linewidth of the signal is only limited by the acquisition time $\tau_{\text{RC}}$ of the lock-in detection. This demonstrates that spectroscopy can be achieved with high resolution, significantly improving upon the method presented in Ref.[11] using pulsed laser excitation that had limited bandwidth and resolution.

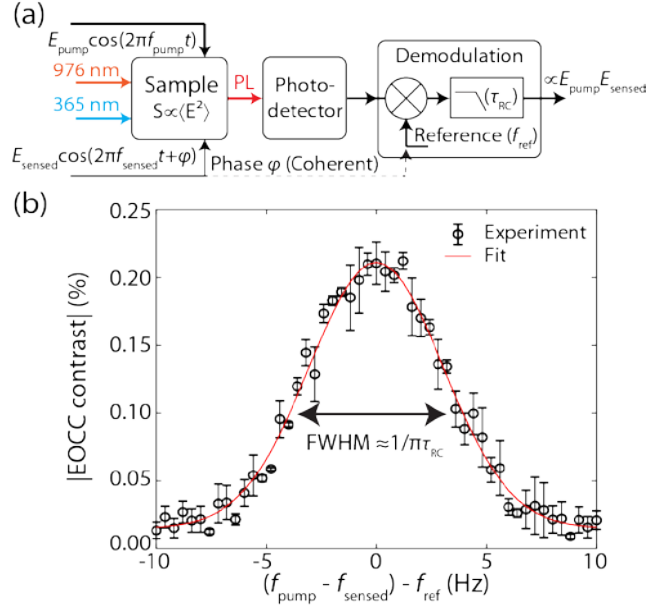

**Fig. 2. Principle of heterodyne electrometry measurements.** (a) Schematic of heterodyne EOCC detection. The sample is illuminated to drive both charge conversion and photoluminescence. A pump electric field ($E_{pump}$) is added on top of the electric field to sense ($E_{sensed}$). The output photoluminescence is measured and demodulated at the reference frequency $f_{ref}$. When the relation $f_{pump} = f_{sensed} + f_{ref}$ is achieved, the output is a product of the two electric fields. If the phase $\varphi$ is known, the detection is coherent. (b) EOCC contrast magnitude as a function of the sensed frequency $f_{sensed}$ measured using incoherent heterodyne detection. The frequency axis is centered to show the difference between the demodulation frequency $f_{ref}$ from the lock-in detector and the beat frequency $f_{pump} - f_{sensed}$. The full-width half-maximum (FWHM) of the response (Gaussian fit) is consistent with the integration time ($\tau_{RC} = 50\ ms$) of the lock-in detector. The experiment is realized at 20 K with the following parameters: $f_{ref} = 397\ Hz$, $E_{pump} = 450\ V/cm$, $f_{pump} = 68\ MHz$ and $E_{sensed} = 55\ V/cm$. Error bars are one standard deviation from the measurement.

We then characterize the EOCC contrast in the center of the device ($X = Y = 0$ µm) as a function of the applied sensed electric field for different $E_{\text{pump}}$ values as shown in Fig. 3(a). The first result is that the signal is now directly proportional to the electric field as expected from heterodyne detection. This is a crucial improvement for sensing applications as small $E_{\text{sensed}}$ values become measurable in contrast with the non-heterodyne quadratic response. The second important result here is the gain provided by $E_{\text{pump}}$ that modifies the intensity of the EOCC contrast. This gain enables both higher sensitivity and control over the dynamic range of the defect sensor. In Fig. 3(b), we compare standard EOCC with both incoherent and coherent heterodyne detection. For the latter, the phase $\varphi$ is fixed which allows to obtain the magnitude after averaging of the in-phase and quadrature components of the lock-in demodulation. The coherent measurement offers a drastic improvement, detecting electric field-induced contrasts lower than the noise baseline obtained in the incoherent case. Overall, we estimate from the experimental acquisition time, signal and noise intensity that our best sensitivity here is 1.1 (V/cm)/√Hz for $E_{\text{pump}} = 750$ V/cm and a near diffraction-limited laser spot size (corresponding to an ensemble size of approximately $10^3$-$10^4$ defects). This sensitivity can be compared with spin-based measurement, where sensitivity of 200 (V/cm)/√Hz for a single defect was reached[18], or 0.1 (V/cm)/√Hz for a very large ensemble ($\sim 10^{11}$ defects, > 50 µm resolution)[19]. Since sensitivity is proportional to the square root of the number of defects (shot-noise limit), our charge-based technique is slightly more sensitive per defect than for single spins and more than two orders of magnitude better for the ensemble.

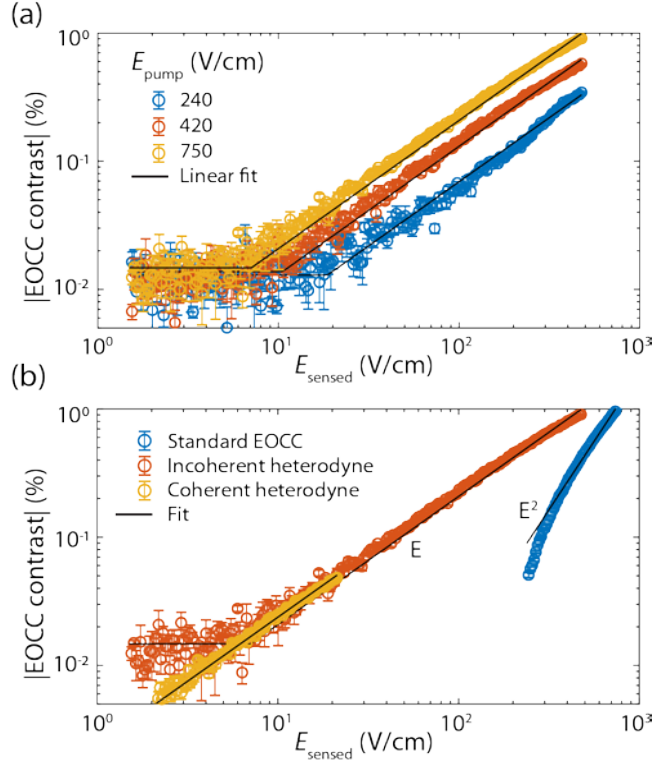

**Fig. 3. Heterodyne responses to electric fields.** (a) Measured EOCC contrast magnitude as a function of the sensed electric field ($E_{sensed}$), and for different pump electric fields ($E_{pump}$) using incoherent (random phase) heterodyne detection. The contrast is found to be linear in the sensed electric field (black line fit), with the slope offset (or gain) increasing with pump amplitude. The observed baseline at low fields is simply the magnitude of the noise. (b) Comparison between the various EOCC methods: standard (direct response without pump)[11], incoherent heterodyne and coherent heterodyne ($E_{pump} = 750\ V/cm$) with lower noise floor. The noise floor is limited by the acquisition time. The fit (black line) is quadratic for standard EOCC and linear for heterodyne detection. In both (a) and (b), the experiment is realized at 20 K with the following parameters: $f_{ref} = 397\ Hz$, $f_{pump} = 68\ MHz$ and $f_{sensed} = f_{pump} + f_{ref}$. Error bars are one standard deviation from the measurement.

The previous experiments were all realized with collinear pump and sensed electric fields, however the heterodyne signal is proportional to the dot product $\vec{u_p} \cdot \vec{u_s}$, such that no oscillating component exists when the two fields are orthogonal. This relationship allows for detection of the sensed field orientation, and in the case of coherent heterodyne, full vector sensing assuming the pump field can be applied both in-plane and out-of-plane. In Fig. 4., we consider the two cases of pump and sensed electric fields applied either parallel or orthogonal as illustrated by the arrows. Both magnitude and phase are obtained directly from the lock-in detection. In the center of the device ($X = Y = 0$ μm) where the protocol is optimized, we observe a significant EOCC magnitude signal in the parallel configuration, whereas the intensity drops to the noise floor in the orthogonal configuration. In the latter case, the phase map has four quadrants in space (excluding near the gates), two for 0 and two for $\pm\pi$, corresponding respectively to aligned and anti-aligned pump and sensed electric fields. This therefore confirms the ability for in-plane vector information using coherent heterodyne. The map becomes more complex near the metallic gates: where the pump field is applied (shaded area, $Y \approx 0$ μm, $|X| \gtrsim 10$ μm), the electric field is strong resulting in saturation and unreliable EOCC response. Where the sensed field is applied ($X \approx 0$ μm, $|Y| \gtrsim 10$ μm), the phase behavior corresponds to the sensed field reorienting from in-plane to out-of-plane with sign flips with respect to the pump. In practice, the sensed field would be obtained only at the center of the device from the lock-in signals in the two pump configurations, $S_\parallel = |A_\parallel| \exp(-i\varphi_\parallel)$ and $S_\perp = |A_\perp| \exp(-i\varphi_\perp)$ with $A$ and $\varphi$ the lock-in magnitude and phase, such that the reconstructed vector signal is $S = S_\parallel + iS_\perp$ in complex space.

Here, out-of-plane information may be obtained from the pump near the metal contacts, however a better device design could involve a back-gate to apply an out-of-plane electric field, e.g. using a bulk N-type substrate beneath the semi-insulating epitaxial layer.

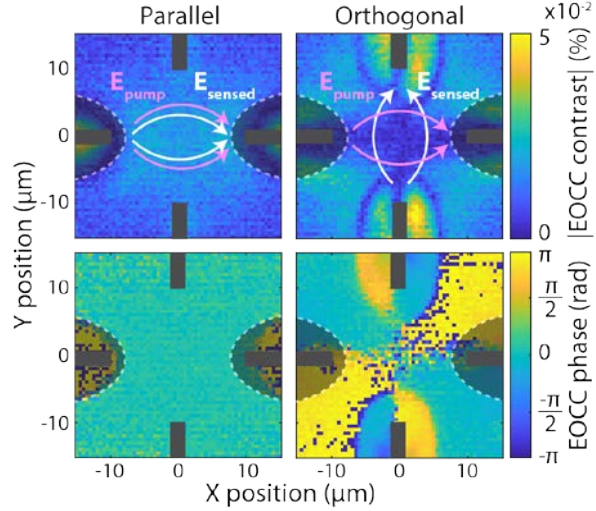

**Fig. 4. In-plane vector sensing.** Pump (pink) and sensed (white) electric fields are applied either parallel (left) or orthogonal (right) as shown by the arrows. Using coherent heterodyne, the EOCC contrast magnitude (top) and phase (bottom) are measured. The magnitude in the center of the device is $2\times 10^{-4}$ and $4 \times 10^{-5}$ (noise floor) for parallel and orthogonal configurations respectively. The phase map is pixelated due to $2\pi$ jumps from noise. The black rectangles represent the metal electrodes, and the shaded area with dashed white border represent saturation behavior from the pump. The experiment is realized at 20 K with the following parameters: $f_{ref} = 397\ Hz$, $f_{pump} = 68\ MHz$, $E_{pump} = 750\ V/cm$, $f_{sensed} = f_{probe} + f_{ref}$ and $E_{sensed} = 7\ V/cm$. The electric field values are estimated for the center of the device.

In conclusion, we show how to apply heterodyne techniques to charge-based defect sensing of radio-frequency electric fields. By tuning the pump field, the sensitivity was enhanced beyond that of spin-based sensing using defects, while enabling a range of capabilities such as high-resolution spectroscopy and vector sensing. Additional improvements could be realized using multi-frequency pumps to obtain simultaneous information over different electric field frequencies, or even of each axis of the vector sensing. Pushing the technique sensitivity beyond the current limit will require a better understanding of the EOCC mechanism, advances in the materials science and devices including denser and larger ensembles. Alternatively, discovering defects tuned for EOCC or working at the single defect limit could provide novel physics to explore and enhanced spatial resolution. Heterodyne detection is well suited for sensing using defects as an external probe, in a scanning microscopy setup for example, and without requiring any external magnetic field. Combined with spin-based magnetic field sensing, defects in solid-state host materials have therefore become remarkable sensors of the local electromagnetic environment.

We thank F. Joseph Heremans and Samuel Bayliss for careful reading of the manuscript. G.W. was supported by the University of Chicago/Advanced Institute for Materials Research (AIMR) Joint Research Center, C.P.A. was supported by the AFOSR was supported by AFOSR FA9550-14-1-0231, S.J.W. was supported by the NSF GRFP, and D.D.A. was supported by the DOE, Office of Basic Energy Sciences. This work made use of the UChicago MRSEC (NSF DMR-1420709) and Pritzker Nanofabrication Facility, which receives support from the SHyNE, a node of the NSF's National Nanotechnology Coordinated Infrastructure (NSF ECCS-1542205).